# This is how the Neocortex Learns

Randall C. O'Reilly

Astera Institute; Department of Psychology
Center for Neuroscience, University of California Davis
correspondence: oreilly@ucdavis.edu

**Abstract:**

A sufficient account of how the neocortex learns must meet three criteria: 1. Computationally, it must approximate a powerful, general-purpose learning algorithm known to scale to human-level intelligence; 2. Algorithmically, it must be implementable using known, well-established neural circuits within the neocortex and associated brain structures; 3. Implementationally, there must be a detailed account for how all of the algorithmic mechanisms actually function at a neurochemical level. At present, there is only one framework that meets all of these criteria: error-driven predictive learning via temporal derivatives, driven by corticothalamic circuits, based on competitive kinase synaptic plasticity induction mechanisms. This has been implemented in the Axon neural simulation framework using spiking neurons, and demonstrated to learn across a wide range of challenging cognitively motivated tasks.

## Introduction

Understanding how the neocortex learns is perhaps the single most important step in understanding human intelligence, because our cognitive functions emerge over years of experience-driven learning within this brain structure, which is unique to mammals and is most greatly expanded in primates, especially humans. Following Marr (1982), there are three levels at which any theory of neocortical learning can be evaluated: *computational, algorithmic,* and *implementational.* The deliberately provocative conclusion of this paper is that only one current theory, which we label the **temporal derivative** model (Jang et al., 2026), provides a sufficient account across these three levels, and thus represents the most viable working hypothesis for how the neocortex learns.

### Computational

Is there a mathematically proven basis for understanding why the neocortical learning mechanism should be capable in principle of accomplishing human-level intelligence through experience-driven learning? Have implementations of this learning mechanism actually demonstrated this capacity? There is only one learning mechanism that meets this criterion: **error backpropagation** (Rumelhart et al., 1986; Widrow & Hoff, 1960; Werbos, 1974), which drives learning in modern-day large-language models (LLMs) based on the transformer architecture (Vaswani et al., 2017), along with essentially every other type of modern powerful neural network model. This is essentially an open-and-shut case at this point: nothing else even comes close.

There have been many proposals for ways of augmenting, simplifying, or approximating error backpropagation, but in practice, nothing else has managed to dethrone the stochastic gradient descent (SGD) procedure, which remains the effectively universal standard for current state-of-the-art (SOTA) models. Furthermore, gradient descent can also be applied to more complex probabilistic models through the *variational inference* approach (Blei et al., 2017), reinforcing the idea that this is perhaps a uniquely efficient way to search through combinatorially large parameter spaces. Recent work provides a more general overarching framework for understanding all of these gradient-based learning mechanisms (Khan & Rue, 2023; Vastola et al., 2026).

Since the skeptical publication of Crick (1989), there has been a widespread and persistent belief among many scientists that error backpropagation is fundamentally incompatible with the biology of the brain. However, in fact there are various longstanding proposals that show how the brain could accomplish backpropagation, which we discuss next.

## Algorithmic

Can the computational-level mechanism be implemented algorithmically in a way that is in principle compatible with known neurobiological mechanisms, at the circuit and structural level? There have been various proposals for how to implement error backpropagation using known neocortical circuits (see Lillicrap et al., 2020 for a recent review, and Discussion below for more details), but only one provides a comprehensive fit with known neocortical (and thalamocortical) properties, while also satisfying the final implementational criterion described next.

The *temporal derivative* algorithm is based on the representation of the backpropagated error gradient *implicitly* as the temporal derivative (difference) between two distinct activation states that emerge over time (Figure 1), instead of requiring a distinct population of neurons to *explicitly* represent the gradient in terms of their firing rates. The idea of using two different states or *phases* of neural activity to compute error gradients originated with the *Boltzmann Machine* (Ackley et al., 1985), and a direct derivation from the error backpropagation formalism was first provided by the *GeneRec* (*Generalized Recirculation*) algorithm (O'Reilly, 1996), which generalized the original *Recirculation* model of Hinton & McClelland (1988). Other derivations were subsequently developed (Xie & Seung, 2003; Scellier & Bengio, 2017).

By representing the error gradient implicitly as a difference between two activation states, this algorithm avoids the significant additional complexity required for maintaining two distinct populations of neurons coding for the feedforward activity and the backpropagated error signal, with a third distinct population representing their difference. Instead, under this temporal derivative framework, all of the neurons in the neocortex are always encoding a positive, mutually compatible representation of the *current* state, through bidirectional excitatory connections that are a well-established and relatively unique property of the neocortex (Markov et al., 2013; Van Essen & Maunsell, 1983). In one phase, this current state reflects a **prediction**, which is then followed by an **outcome** that may differ from the prior prediction, with any difference in corresponding activation states throughout the network providing a close approximation to the error gradient between outcome and prediction.

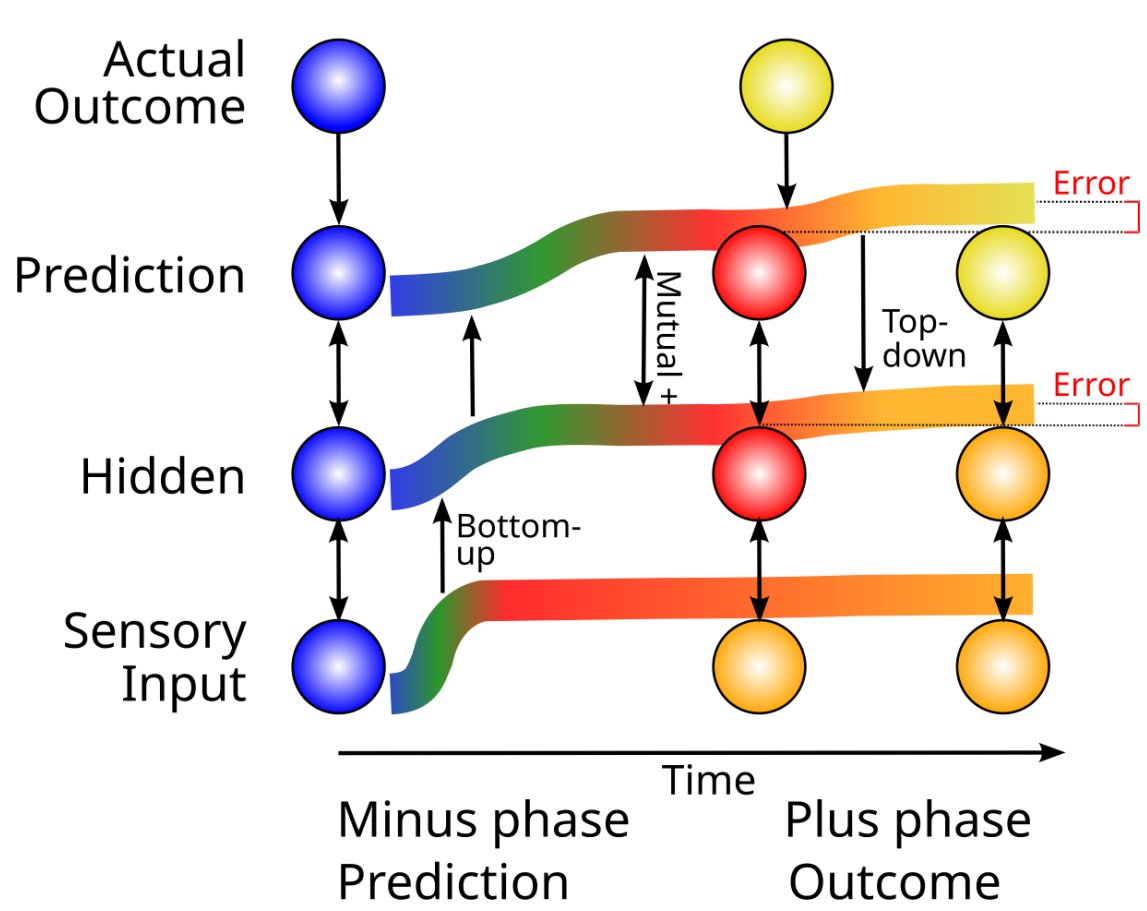

**Figure 1:** How bidirectional activation propagation can communicate error signals, in the simplest case of a three-layer network mapping from a Sensory Input to a Prediction output, with the Actual Outcome driving the Prediction layer only in the later *plus* phase, after an initial *minus* phase when the prediction is generated. The Error is the temporal difference between the (*plus* – *minus*) activity levels. There is just one network with three bidirectionally connected units, as shown at the left; the networks shown further to the right are snapshots of the activity state of this network at different points in time, which evolves from left to right. The thick colored lines also show the activation level of each of the three neurons over time, both in terms of the line height and the brightness and warmth of the color gradient. Initially, each neuron is inactive (blue). Then, external Sensory Input arrives, and a wave of bottom-up excitation propagates upward through the Hidden and Prediction layers. Critically, the Prediction and Hidden neurons mutually excite each other via bidirectional connections, which contributes to each of their activity levels. The snapshot of the network in the middle shows the neural activity at the end of the minus phase. Then, at the start of the plus phase, the Actual Outcome arrives, which is more active than the Prediction, and it therefore drives more activity in the Prediction neuron. This propagates top-down to the Hidden neuron as well, which is the key mechanism by which bidirectional connectivity communicates error signals, causing the Hidden neuron to have a (*plus* – *minus*) activity difference, reflecting the top-down influence from the Prediction layer. This temporal-difference based Error signal provides a good approximation to the error backpropagation error gradient (O'Reilly, 1996).

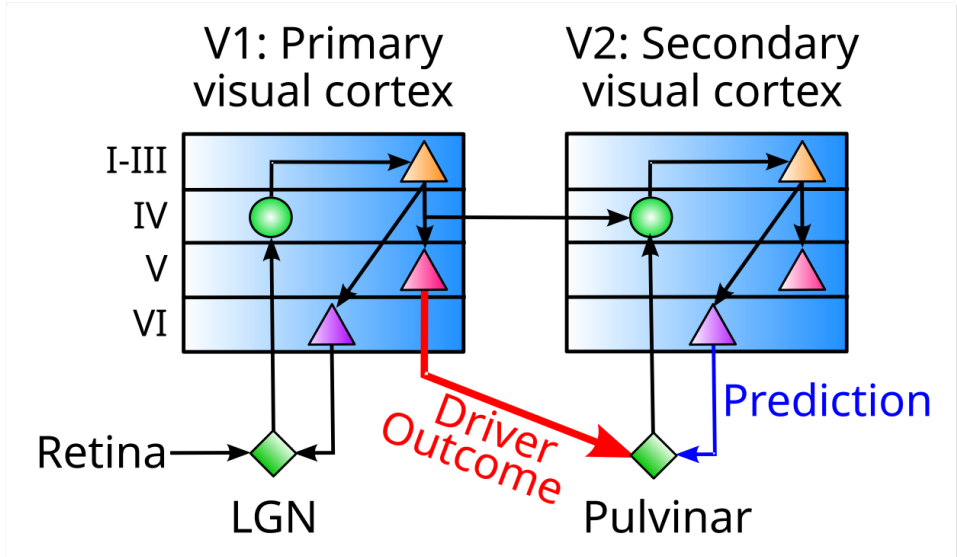

**Figure 2:** Connectivity between the neocortex and the pulvinar nucleus of the thalamus, in the case of primary and secondary visual areas, that is uniquely well-suited for driving predictive error-driven learning. The numerous and relatively weaker projections from layer 6 (VI) neurons activate a prediction over the pulvinar, that integrates the signals from multiple cortical areas and neurons to synthesize the prediction, which improves over the course of learning throughout the neocortex and in these final projections into the pulvinar. By contrast, the strong, focal driver inputs from layer 5 (V) intrinsic bursting (5IB) neurons can activate an outcome representation that is essentially an unlearned copy of the activity pattern in lower cortical layers (e.g., V1 trains V2 predictions in this case). The periodic bursting of the 5IB neurons ensures that this outcome activity is only phasically present (i.e., the plus phase), with a complete prediction – outcome learning cycle occurring within roughly 200 ms (i.e., theta frequency, 5 Hz). Diagram based on Sherman & Guillery (2006).

The thalamocortical connectivity of the neocortex directly supports the generation of these two states, within an overall **predictive learning** framework, which is learning to predict what happens next based on what just happened (O'Reilly et al., 2021). Specifically, thalamic relay cells (TRCs) in the higher-order *pulvinar* (for posterior cortex) and *mediodorsal* (for frontal cortex) nuclei of the thalamus receive two distinct inputs (Figure 2; Sherman & Guillery, 2006; Usrey & Sherman, 2018):

1. A large number of normal-strength inputs onto more distal dendritic arbors from a wide range of higher-level neocortical brain areas.

2. A much smaller number (even just one) of abnormally strong *driver* inputs originating from the layer 5b intrinsic bursting (5IB) neurons in hierarchically lower neocortical areas, which burst roughly every 100-200 ms (i.e., at the alpha or theta rhythm), and are otherwise inactive.

This constellation of properties creates phasic alternations between a prediction state driven exclusively by the first pathway, and an outcome state that reflects the strong impact of the driver inputs, which only occurs phasically. Critically, the pulvinar neurons *do not* compute the difference between these two inputs and send that as their output, instead they merely *reflect* these two different inputs in their activity levels *across time.*

The TRCs send extensive excitatory reciprocal projections back to the same areas that send the prediction-generating inputs, thereby communicating the temporal difference between the prediction and outcome states back into the neocortex, where local synaptic plasticity is driven by this temporal derivative. Furthermore, bidirectional connectivity within the neocortex effectively computes the partial derivative of these error signals relative to sending neural activity coming from other areas, thereby accomplishing the critical error backpropagation *credit assignment* learning process between neocortical layers (O'Reilly, 1996).

There is a wealth of detailed neuroscience data that is consistent with this overall framework, as reviewed in O'Reilly et al. (2021) (e.g., Fiebelkorn & Kastner, 2021; Sherman & Guillery, 2006; Sherman & Usrey, 2024).

## Implementational

Going down a level of neurobiological detail, how could the algorithmic property of a temporal derivative between prediction and outcome states actually drive synaptic plasticity locally at all of the neocortical and thalamic synapses? Mathematically, the temporal derivative can be computed as the difference between *fast* minus *slow* integrals of a common driving input signal. Intuitively, the fast integral more closely reflects the more recent outcome state, while the slow integral still retains more of the trace from the earlier prediction state. See temporal derivative on compcogneuro.org for an interactive demonstration of this principle.

Neurochemically, the difference between LTP (long term potentiation, i.e., synaptic weight increase) versus LTD (long term depression, weight decrease) is determined in part by a competition between two different *kinases*, *CaMKII* (calcium calmodulin kinase II) and *DAPK1* (death-associated protein kinase 1), both of which are driven by calcium-activated calmodulin (CaM) (Goodell et al., 2017; Goodell et al., 2021; Cook et al., 2021; Tullis & Bayer, 2023; Bayer & Giese, 2025). If CaMKII has a faster overall integration of the common CaM driver, and DAPK1 a slower such integration, then this would implement the necessary temporal derivative mechanism.

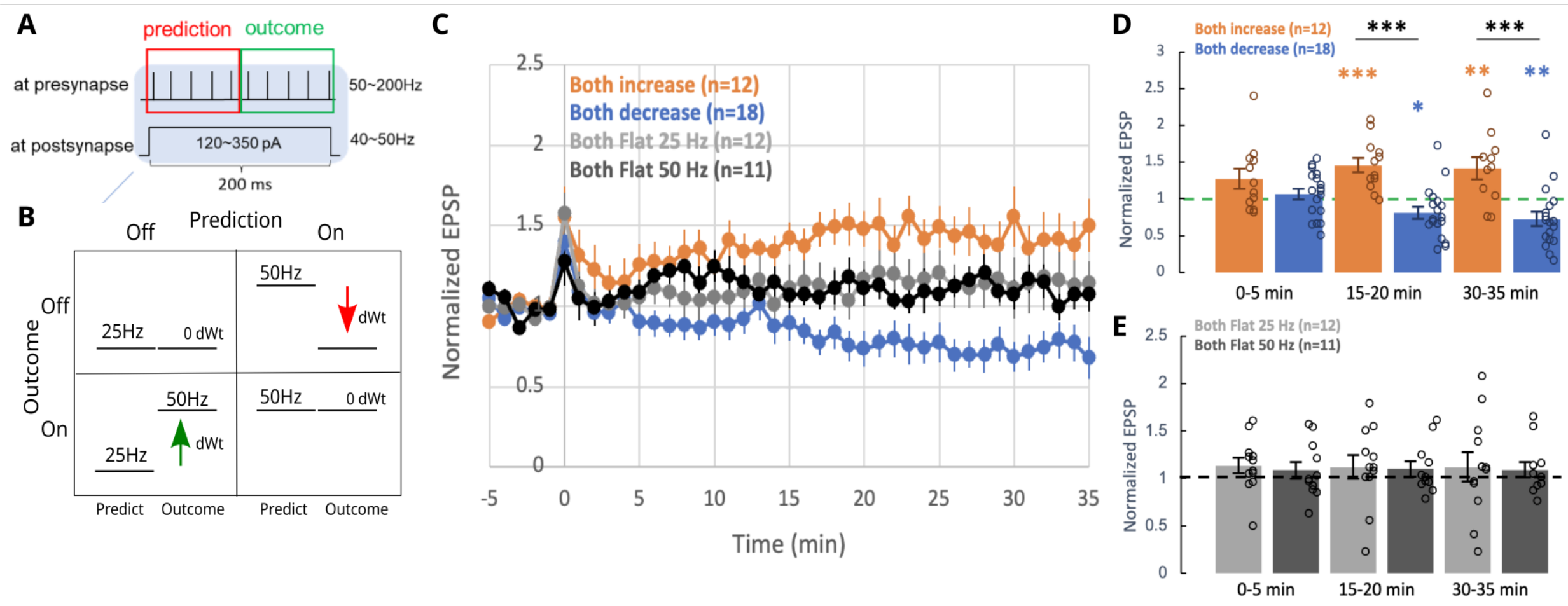

**Figure 3:** Results from Jang et al. (2026), which are consistent with the predictions of the temporal derivative learning mechanism. **A** Pre- and postsynaptic neurons were stimulated to either 25 Hz or 50 Hz across the two 100 ms halves (prediction, outcome) of a 200 ms theta cycle. **B** All 4 cells of the 2x2 combination of prediction, outcome frequencies were tested. **C** The progression of probe EPSP amplitudes surrounding the stimulation protocol at time 0, showing that the increasing temporal derivative (25 to 50 Hz for both the pre and postsynaptic neurons, in orange) resulted in LTP, while the decreasing temporal derivative (50 to 25 Hz, in blue) resulted in LTD. Both flat profiles (constant 25 Hz or 50 Hz) resulted in no net synaptic efficacy change. **D**, **E** Summary showing all cell values (open circles) and averages (bars) for the individual conditions at different times after stimulation, with statistically significant results highlighted with asterisks (** = P < .01, *** = P < .001).

There is now direct experimental evidence consistent with this prediction (Jang et al., 2026). Specifically, pre and postsynaptic pyramidal neurons in a widely-used *in vitro* synaptic plasticity preparation were driven by different temporal patterns of activity over a 200 ms theta-cycle window, with one activity level for the first 100 ms (i.e., reflecting the prediction state), and another activity level for the second 100 ms (the outcome state).

After 10 repeated presentations of these different temporal patterns, the resulting changes from baseline synaptic efficacy strength matched the predictions of the temporal derivative learning mechanism (Figure 3). Specifically, when there was a rising pattern between prediction and outcome (25 Hz to 50 Hz), LTP resulted. When this pattern went the opposite direction (50 Hz to 25 Hz), LTD resulted. Finally, and critically, for *both* of the stable conditions (25 Hz to 25 Hz and 50 Hz to 5 Hz), no net synaptic efficacy change occurred. This latter condition directly contradicts the standard Hebbian-style account of synaptic plasticity, because the 50-50 case has the most overall synaptic activity, and yet it did not result in LTP, whereas the 25-to-50 case did. Furthermore the 25-to-50 and 50-to-25 cases both have the same net amount of synaptic activity, just organized differently across time.

### Summary

Thus, the strong conclusion from this summary evaluation is that the temporal derivative form of error-driven predictive learning is unique in providing a consistent and empirically supported account for how the neocortex learns, across all three relevant levels of analysis. Furthermore, this theory has been implemented in large-scale spiking neural networks, in the Axon framework, described extensively at compcogneuro.org. That website provides many examples of these models that can be run through the web browser, using the WebGPU framework for GPU-based acceleration to provide reasonable performance. For further details, see the kinase algorithm.

## Alternative frameworks

Various points of contrast with other possible learning algorithms are briefly discussed below, to clarify the relevant distinctions.

### Explicit error and predictive coding

Various alternative proposals for implementing the error backpropagation algorithm (Lillicrap et al., 2020) and the classic Bayesian predictive coding framework (e.g., Rao & Ballard, 1999) both hypothesize that a sub-population of neurons directly represent the *error,* by subtracting a top-down prediction from the bottom-up actual outcome (Figure 4). Thus, different populations of neurons must be somehow segregated so that they can represent fundamentally distinct information. Furthermore, all three of these different signals (prediction, outcome, error) should in principle be communicated across layers, in different directions, requiring strongly segregated pathways.

By contrast, as emphasized above, the temporal derivative framework keeps the error gradient representation *implicit*, as the difference in activity states over time, which greatly simplifies the biological implementation required. Furthermore, it allows for all levels in the network to work together to drive parallel *constraint satisfaction* processing, integrating top-down and bottom-up constraints, to drive coherent interpretations of the current state (Hopfield & Tank, 1985; O’Reilly et al., 2013). This represents a powerful form of search through representation space, operating as a kind of inner-loop optimization within the outer-loop of error backpropagation search through synaptic weight space to improve the predictive accuracy of the system.

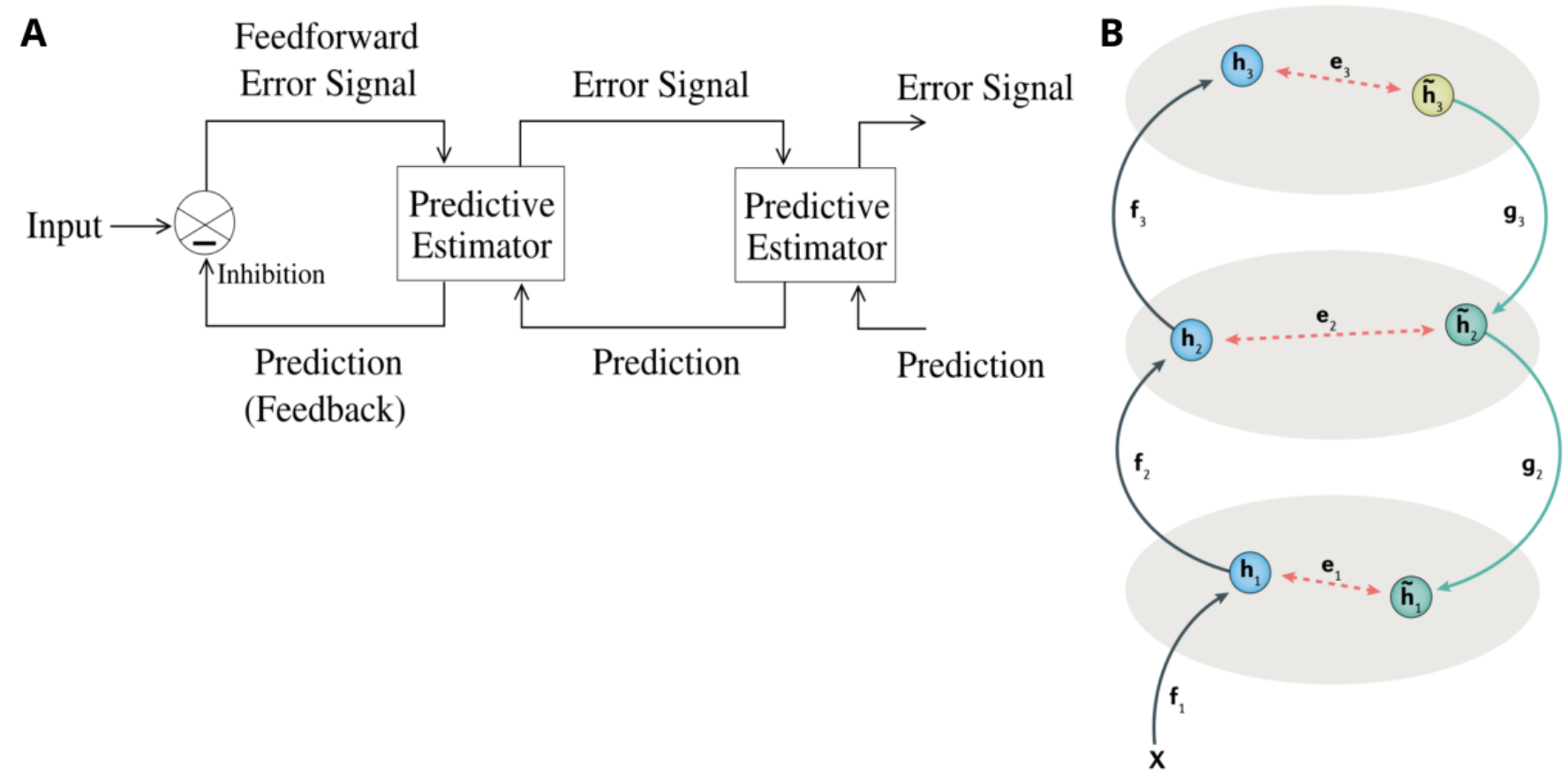

**Figure 4:** Proposals for implementing error-driven learning that require explicit error signals and / or separate neural pathways for feedforward vs. prediction signals. Given the pervasive interconnectivity of all lamina and neurons in the neocortex, it is unlikely that such strict separation between such channels is sustained. **A** is from Rao & Ballard (1999) on predictive coding, where the error is explicitly represented in neural firing, as a difference between a top-down prediction and bottom-up signal. Note that the feedforward signal to higher layers is *exclusively* an error signal, not a positive representation of the input stimulus. **B** is the target-prop model from Le Cun (1986) (via Lillicrap et al., 2020), which has a separate feedforward pathway on the left, and top-down feedback on the right, with the difference between these two providing the error signal.

The available neural evidence is consistent with the coherent, synergistic, redundant encoding of information across all levels of the cortex, with no significant evidence of the kind of structural segregation required by the explicit error models (Walsh et al., 2020; Heilbron & Chait, 2018). Furthermore, the standard predictive coding framework only has error signals propagating forward beyond the first layer, which is inconsistent with the extensive evidence showing that higher cortical areas contain positive representations of the input stimulus, at various levels of abstraction.

Thus, the temporal derivative framework supports the widely accepted idea that the neocortex learns by generating top-down predictions of what will happen next, in a way that appears to be more compatible with available neural evidence at multiple levels of analysis.

## Hebbian learning

The predominant computational-level interpretation of neocortical learning in the literature has generally focused on various forms of Hebbian learning, based on the well-established data demonstrating a relationship between the level of postsynaptic calcium, entering via NMDA receptors, and the direction and magnitude of synaptic plasticity (Lisman, 1989; Bear & Malenka, 1994). Specifically, low levels of calcium result in LTD, while higher levels result in LTP. This is generally consistent with the BCM (Bienenstock et al., 1982) version of a Hebbian learning algorithm.

Despite this seeming advantage at the implementational level, Hebbian learning is essentially a non-starter at the computational level, because it only has a local, heuristic function in terms of extracting statistical regularities of co-activation (Oja, 1982; Rumelhart & Zipser, 1985; Intrator & Cooper, 1992). Therefore, there is no reason to believe that Hebbian learning can effectively train deep layered networks like those present in the neocortex, whereas this is precisely the case where error backpropagation excels.

The spike-timing dependent plasticity (STDP) (Bi & Poo, 1998) version of Hebbian learning has been a primary focus of computational models (e.g., Kheradpisheh et al., 2018; Diehl & Cook, 2015). However, it is now clear that the simple computationally compelling form of STDP originally described, which required a very particular stimulation protocol with individual pairs of spikes separated by 1 s intervals, is not generally applicable to more realistic patterns of neural activity (Debanne & Inglebert, 2023). Indeed the same BCM-like pattern emerges with more realistic, denser activity patterns (Shouval et al., 2010; Izhikevich & Desai, 2003). Thus, STDP lacks both implementational-level support and a coherent computational-level account for why it should be a powerful learning mechanism.

## Eligibility traces and specialized output learning

There is increasing evidence for a form of learning that applies specifically to output neurons in the neocortex and the hippocampus, which may function somewhat like the output decoding neurons in *reservoir computing* networks (Verstraeten et al., 2007; Tanaka et al., 2019), where a complex internal dynamical state can be read out by only adapting a single layer of output neuron synapses. This has been termed *behavioral timescale synaptic plasticity (BTSP)*, and extensively studied in area CA1, which is the output layer of the hippocampus (Magee, 2026; Bittner et al., 2015; Bittner et al., 2017).

In BTSP, elevated *plateau potentials* in distal dendrites provide the critical plasticity-inducing mechanism that establishes an *eligibility trace* that lasts for several seconds. In CA1, these distal plateau potentials are activated by entorhinal cortex layer 3 inputs, which thus serve as a special training signal for driving plasticity in the other major population of synaptic inputs, from area CA3. A similar mechanism has recently been described in layer 5 pyramidal neurons in neocortex, which also have a prominent distal dendritic tuft, and are the primary output neurons of the neocortex (Yaeger et al., 2025; Xiao et al., 2025).

In both of these BTSP cases, there is rapid learning driven by the distal dendritic inputs, with an inhibitory negative feedback loop that prevents overtraining (Campbell et al., 2026). There is evidence in the CA1 neurons that this plasticity is generally transient, consistent with a *fast mapping* type of learning that can rapidly adapt to read out behaviorally-relevant signals from the more slowly-adapting internal representations of the relevant systems (hippocampus or neocortex) (Vaidya et al., 2025).

The driving target signal for plasticity in the layer 5 neocortical neurons remains unclear (Magee, 2026), but we do know that these neurons receive extensive thalamic input targeting the distal dendritic tuft in layer 1. The thalamic projections that target layer 1 are generally of the *matrix* type, which means they typically have broad axonal arbors targeting many different neurons across multiple cortical areas. A prominent and widespread source of such projections comes from the ventral anterior (VA) nucleus, which receives inputs from layer 5 neurons in motor cortical areas, and is also under disinhibitory control from the basal ganglia (Phillips et al., 2021; Xiao et al., 2009; Kuramoto et al., 2015; Economo et al., 2018). This would provide a way for direct motor-relevant target signals to drive the rapid tuning of neocortical output neurons across most of neocortex, even all the way down in area V1 (Yaeger et al., 2025; Kuramoto et al., 2015).

The temporally-extended nature of the eligibility-trace mechanism allows later outcome or motor action signals to drive learning from earlier state representations, providing a solution to the *temporal* version of the credit assignment process (this is what the behavioral timescale connotes). In other brain areas such as the basal ganglia, and recently in the neocortex, neuromodulatory signals including dopamine have been found to drive eligibility-trace learning mechanisms, bridging the temporal gap until reinforcement signals become available (He et al., 2015; Shouval & Kirkwood, 2025).

The current evidence suggests that the BTSP-based eligibility-trace mechanisms are synergistic with the more slowly-accumulating and shorter time-scale plasticity mechanisms (Magee, 2026), which provide the initial biases and encodings upon which the fast BTSP learning builds. The logic is similar to the complementary learning systems framework for understanding how the hippocampus and neocortex are separately optimized for rapid *episodic* learning (hippocampus) and slow *statistical* learning (neocortex) (McClelland et al., 1995; O'Reilly et al., 2014): error-driven learning within the many layers of the neocortex slowly learns to extract systematic ways of representing the structure of the environment, while the rapid, output-focused BTSP mechanism provides a way to quickly decode the resulting complex internal states to satisfy the current behavioral demands.

## Conclusion

Across the three levels considered here, the strongest constraint appears to come from the computational level, effectively narrowing the field to one viable type of learning: error backpropagation. If someone were to discover something even more generally powerful than error backpropagation, that would certainly represent an important advance across many fields, but given the massive amount of research that has been invested into exploring algorithms at the computational level, this is seeming increasingly unlikely.

This then puts more of the weight on the algorithmic and implementational-level arguments outlined above, which each now have significant empirical evidence to support them. Nevertheless, more extensive empirical research is essential to further test between the different possible algorithmic and implementational possibilities that have been advanced for accomplishing predictive error-driven learning.